\documentstyle[12pt,fleqn,epsfig]{article}

\newlength{\dinwidth}
\newlength{\dinmargin}
\begin{document}
\begin{flushright}
MC-TH-2002/14\\
Cavendish-HEP-2003/01\\
February 2003\\
\end{flushright}
\begin{center}
\vspace*{2cm}

{\Large {\bf Mass bounds in a model with a Triplet Higgs}} \\

\vspace*{1cm}

J.R. Forshaw$^1$, A.~Sabio Vera$^2$ and B.E. White$^1$

\vspace*{0.5cm}
$^1${\sl Department of Physics \& Astronomy},\\
{\sl University of Manchester},\\
{\sl Oxford Road},\\
{\sl Manchester M13 9PL, U.K.}\\\vspace{0.5cm}
$^2$ {\sl Cavendish Laboratory},\\
{\sl University of Cambridge},\\
{\sl Madingley Road},\\
{\sl Cambridge CB3 OHE, U.K.}\\
\end{center}
\vspace*{2cm}

\begin{abstract}
{We perform an analysis of the Renormalization Group evolution 
of the couplings in an extension to the Standard 
Model which contains a real triplet in the Higgs sector. Insisting that
the model remain valid up to 1~TeV allow us to map
out the region of allowed mass for the Higgs bosons. We conclude
that it is possible for there to be no light Higgs bosons without
any otherwise dramatic deviation from the physics of the Standard Model.}

\end{abstract}

\newpage

\section{Introduction}
In a previous paper, we studied an extension of the Standard Model
in which a real scalar $SU(2)$ triplet with zero hypercharge is added
to the usual scalar $SU(2)$ doublet \cite{Forshaw:2001xq}. 
We showed that such an extension
is allowed by the precision data and that the mass of the lightest
Higgs boson can be as big as 500~GeV. 

To recap, the Lagrangian of the model in terms of the 
usual Standard Model Higgs, $\Phi_1$, and the new triplet, $\Phi_2$, reads
\begin{eqnarray}
{\cal L} &=& (D_\mu \Phi_1)^{\dagger}~ D^\mu \Phi_1 
+ \frac{1}{2} (D_\mu \Phi_2)^{\dagger} ~D^\mu \Phi_2 
- V_0(\Phi_1 , \Phi_2),
\label{eqn:lag}
\end{eqnarray}
with a scalar potential
\begin{eqnarray}
V_0(\Phi_1 , \Phi_2) &=&\mu_1^2 ~|\Phi_1|^2 + \frac{\mu_2^2}{2} 
~|\Phi_2|^2 + \lambda_1 ~|\Phi_1|^4+ \frac{\lambda_2}{4} ~|\Phi_2|^4 
+ \frac{\lambda_3}{2} ~|\Phi_1|^2 ~|\Phi_2|^2 \nonumber\\
&+& \lambda_4 \,  {\Phi_1}^\dagger \sigma^\alpha \Phi_1 ~ {\Phi_2}_\alpha.
\end{eqnarray}
$\sigma^\alpha$ are the Pauli matrices.
The expansion of the field components is
\begin{eqnarray}
\Phi_1 = \left(\begin{array}{c}\phi^+ \\
\frac{1}{\sqrt{2}}\left(h_c^0 + h^0 + i \phi^0\right)\end{array}\right)_
{Y=1},
~~\Phi_2 = \left(\begin{array}{c}\eta_1 \\
\eta_2 \\ \eta_c^0 + \eta^0    \end{array}\right)_{Y=0}
\end{eqnarray}
where $\eta^\pm = ( \eta_1 \mp i \eta_2) / \sqrt{2}$ and $\phi^0$ is the
Goldstone boson which is eaten by the $Z^0$.

The model violates custodial symmetry at tree level giving a prediction 
for the $\rho$-parameter of  
\begin{eqnarray}
\rho ~=~ 1 + 4 \left(\frac{\eta^0_c}{h^0_c}\right)^2.
\end{eqnarray}
As discussed in \cite{Forshaw:2001xq}, it is precisely this violation
of custodial symmetry which allows the lightest Higgs to be much 
heavier than in the Standard Model. By giving the triplet a non-zero
vacuum expectation value, one is in effect making a positive 
tree-level contribution to the $T$-parameter, and this is enough to allow
a heavier Higgs.

In the neutral Higgs sector we have two CP-even states which mix 
with angle $\gamma$. The mass eigenstates $\{H^0, N^0\}$ are defined by 
\begin{eqnarray}
\left(\begin{array}{c}H^0\\N^0\end{array}\right) &=& 
\left(\begin{array}{cc}\cos{\gamma}&-\sin{\gamma}\\
\sin{\gamma}&\cos{\gamma}\end{array}\right)
\left(\begin{array}{c}h^0\\\eta^0\end{array}\right).
\end{eqnarray}
There is also mixing in the charged Higgs sector. We define the mass
eigenstates $\{g^\pm, h^\pm\}$ by 
\begin{eqnarray}
\left(\begin{array}{c}g^\pm\\h^\pm\end{array}\right) &=& 
\left(\begin{array}{cc}\cos{\beta}&-\sin{\beta}\\
\sin{\beta}&\cos{\beta}\end{array}\right)
\left(\begin{array}{c}\phi^\pm\\\eta^\pm\end{array}\right).
\end{eqnarray}
The $g^\pm$ are the Goldstone bosons corresponding to $W^\pm$ and,
at tree level, the mixing angle is
\begin{eqnarray}
\tan{\beta} &=& 2 \frac{\eta^0_c}{h^0_c}.
\end{eqnarray}
The precision electroweak data constrain $\beta$ to be smaller
than about $4^\circ$ \cite{Forshaw:2001xq}.

In this paper we wish to examine the renormalization group flow of the
couplings and hence establish bounds on the scalar masses under the
assumption that the triplet model remain valid up to some scale $\Lambda$.
We take $\Lambda = 1$ TeV and make no statements about physics
at higher scales. For the Lagrangian of (\ref{eqn:lag}) to remain
appropriate up to $\Lambda$, we demand that the scalar couplings $\lambda_i$
remain perturbative and that the vacuum remain stable (i.e. is a local minimum)
up to $\Lambda$. We begin in the next section with the calculation of
the beta-functions. In Section 3 we present our results away from the
decoupling limit of the model and in Section 4 we discuss decoupling.  

\section{The one-loop effective potential and the beta-functions}

The effective potential 
\cite{Coleman:jx,Jackiw:cv,Iliopoulos:ur,Kastening:1991,Ford:mv,Quiros:dr} 
has the following one-loop expansion in the 
$\overline{\rm MS}$ renormalization scheme and 't~Hooft-Landau gauge: 
\begin{eqnarray}
\hspace{-2cm}V &=& V_0 + V_{\rm CT} + V_1  \nonumber\\
&=& 
\frac{1}{2}{{{\mu }_1}}^2\,{h^0_c}^2 + 
  \frac{1}{2}{{{\mu }_2}}^2\,{\eta^0_c}^2 + 
  \frac{1}{4}{{\lambda }_1}\,{h^0_c}^4 + 
  \frac{1}{4}{{\lambda }_2}\,{\eta^0_c}^4+ 
  \frac{1}{4}{{\lambda }_3}\, {h^0_c}^2 \, {\eta^0_c}^2- 
  \frac{1}{2}{\lambda_4} \,{h^0_c}^2\,{\eta^0_c} \nonumber\\
&+& \delta \Omega 
-\frac{1}{2} \delta{{{\mu }_1}}^2\,{h^0_c}^2
- \frac{1}{2} \delta{{{\mu }_2}}^2\,{\eta^0_c}^2
+ \frac{1}{4}\delta{{\lambda }_1}\,{h^0_c}^4 + 
  \frac{1}{4}\delta{{\lambda }_2}\,{\eta^0_c}^4
+ \frac{1}{4}\delta{{\lambda }_3}\,{h^0_c}^2\,{\eta^0_c}^2 
- \frac{1}{2}\delta\lambda_4 \,{h^0_c}^2\,{\eta^0_c} \nonumber\\
&+& \frac{1}{16 \pi^2} \left\{
\frac{3}{4} {m}_Z^4 \left(\log{\frac{{m}_Z^2}{\mu^2}}
-\frac{5}{6}\right)
+\frac{3}{2} {m}_W^4 \left(\log{\frac{{m}_W^2}{\mu^2}}
-\frac{5}{6}\right) 
-3~ {m}_t^4 \left(\log{\frac{{m}_t^2}{\mu^2}}
-\frac{3}{2}\right)\right.\nonumber\\
&+&\frac{1}{4} {m}_{\phi^0}^4
\left(\log{\frac{{m}_{\phi^0}^2}{\mu^2}}-\frac{3}{2}\right)
+\frac{1}{2} {m}_{g^\pm}^4 
\left(\log{\frac{{m}_{g^\pm}^2}{\mu^2}}
-\frac{3}{2}\right)+\frac{1}{2} {m}_{h^\pm}^4 
\left(\log{\frac{{m}_{h^\pm}^2}{\mu^2}}-\frac{3}{2}\right)\nonumber\\
&+&\left.\frac{1}{4} {m}_{H^0}^4 \left(\log{
\frac{{m}_{H^0}^2}{\mu^2}}
-\frac{3}{2}\right)+\frac{1}{4} {m}_{N^0}^4 
\left(\log{\frac{{m}_{N^0}^2}{\mu^2}}-\frac{3}{2}\right)\right\} \nonumber\\
&-& \frac{C_{\rm UV}}{64 \pi^2} \left\{
3 ~ {m}_Z^4 +6 ~{m}_W^4 - 12 ~ {m}_t^4 + {m}_{\phi^0}^4
+ 2 ~{m}_{g^\pm}^4 + 2 ~{m}_{h^\pm}^4 + {m}_{H^0}^4 + {m}_{N^0}^4 \right\}.
\end{eqnarray}
$\mu$ is the renormalization scale and 
$C_{\rm UV} = \frac{2}{4-D}-\gamma_E + \log{4 \pi}$.
We have included the contributions from 
all the relevant physical states including the heaviest fermion, the top quark.
The terms with $\delta$ correspond to the counterterms of the theory and 
the tree-level masses are
\begin{eqnarray}
m_Z^2 &=&\frac{1}{4}{h^0_c}^2\,\left( g^2 + {g'}^2 \right), \\
m_W^2 &=& \frac{1}{4}g^2\,{h^0_c}^2 + g^2\,{\eta^0_c}^2, \\
m_t^2 &=& \frac{1}{2}{h_t}^2\,{h^0_c}^2, \\
m_{\phi^0}^2 &=& {{{\mu }_1}}^2 + {{\lambda }_1}\,{h^0_c}^2 + 
  \frac{1}{2} \, \lambda_3 \, {\eta^0_c}^2 - \lambda_4 \, \eta^0_c,\\
m_{g^\pm}^2 &=& \mu_1^2 + \lambda_1 \, {h^0_c}^2 + \lambda_4 \, \eta^0_c 
+ \frac{1}{2}\,\lambda_3\,{\eta^0_c}^2 - \lambda_4 \, h^0_c \, \tan{\beta},\\
m_{h^\pm}^2 &=& \mu_2^2 + \lambda_2 \, {\eta^0_c}^2 + \lambda_4 \, h^0_c 
\, \tan{\beta} + \frac{1}{2}\,\lambda_3\,{h^0_c}^2,\\ 
m_{H^0}^2 &=& {{{\mu }_1}}^2 + 3\,{{\lambda }_1}\,{h^0_c}^2 + 
  \frac{1}{2}\,\lambda_3 \,{\eta^0_c}^2 - \lambda_4 \, \eta^0_c 
+ \lambda_4 \, h^0_c \, \tan{\gamma} 
- \lambda_3 \, h^0_c \, \eta^0_c \, \tan{\gamma}, \\
m_{N^0}^2 &=& \mu_2^2 + 3\, \lambda_2 \, {\eta^0_c}^2 - \lambda_4 \, h^0_c \, 
\tan{\gamma} + \frac{1}{2}\, \lambda_3 \, h^0_c \left(h^0_c + 2 \, 
\eta^0_c \, \tan{\gamma}\right).
\end{eqnarray}
It is understood that we should substitute explicitly for the mixing
angles, which are solutions to the equations
\begin{eqnarray}
&&\hspace{-1cm}\lambda_4 h^0_c + \tan{\beta}\left(\mu_1^2 - \mu_2^2 +
\lambda_1 {h^0_c}^2
- \frac{1}{2} \lambda_3 {h^0_c}^2 + \lambda_4 \eta^0_c - \lambda_2
{\eta^0_c}^2 + \frac{1}{2}\lambda_3 {\eta^0_c}^2 - \lambda_4 h^0_c
\tan{\beta}\right) = 0,\label{eq:beta} \\
&&\hspace{-1cm}-\lambda_4 h^0_c + \lambda_3 h^0_c \eta^0_c +
\tan{\gamma} \left(\mu_1^2 - \mu_2^2 + 3 \lambda_1 {h^0_c}^2
-\frac{1}{2}\lambda_3 {h^0_c}^2 - \lambda_4 \eta^0_c \right. \nonumber\\
&&\left. - 3 \lambda_2
{\eta^0_c}^2 + \frac{1}{2} \lambda_3 {\eta^0_c}^2 + \lambda_4 h^0_c
\tan{\gamma} - \lambda_3 h^0_c \eta^0_c \tan{\gamma}\right) = 0. 
\label{eq:gamma}
\end{eqnarray}

The expressions for the counterterms are thus 
\begin{eqnarray}
\delta \Omega &=& \frac{C_{\rm UV}}{64 \pi^2}\,
\left(4\,\mu_1^4 + 3\,\mu_2^2\right),\\
\delta \mu_1^2 &=& -\frac{C_{\rm UV}}{32 \pi^2}\,\left(12\,\lambda_1\,\mu_1^2 
+ 3 \lambda_3 \, \mu_2^2 
+ 6 \, \lambda_4^2 \right),\\
\delta \mu_2^2 &=& -\frac{C_{\rm UV}}{32 \pi^2}\,\left(10\,\lambda_2\,\mu_2^2 
+ 4 \, \lambda_3 \mu_1^2 
+ 4 \, \lambda_4^2\right),\\
\delta \lambda_1 &=& \frac{C_{\rm UV}}{16 \pi^2}\,\left(\frac{9}{16}\,g^4 
- 3\, h_t^4 + 12\,\lambda_1^2
+\frac{3}{4}\,\lambda_3^2 + \frac{3}{8}\,g^2\,{g'}^2+\frac{3}{16}\,{g'}^4
\right),\\
\delta \lambda_2 &=& \frac{C_{\rm UV}}{16 \pi^2}\,\left(6\,g^4 + 11\,
\lambda_2^2 + \lambda_3^2\right),\\
\delta \lambda_3 &=&  \frac{C_{\rm UV}}{16 \pi^2}\,
\left(3\,g^4 + 6\,\lambda_1 \, \lambda_3 
+5 \, \lambda_2 \, \lambda_3 + 2 \, \lambda_3^2\right),\\
\delta \lambda_4 &=&  \frac{C_{\rm UV}}{8 \pi^2}\,\lambda_4 \, 
\left(\lambda_1 + \lambda_3 \right),
\end{eqnarray}
where $\delta \Omega$ is the counterterm for the vacuum energy.

The fact that the theory should be independent of the unphysical mass $\mu$ 
implies that the couplings and masses acquire a $\mu$ dependence governed by 
the Renormalization Group (RG) equation for the one-loop effective 
potential, i.e.
\begin{eqnarray}
&&\left(\beta_{\mu_1} \frac{\partial}{\partial \mu_1^2}+ 
\beta_{\mu_2} \frac{\partial}{\partial \mu_2^2}+ 
\beta_{\lambda_1} \frac{\partial}{\partial \lambda_1}+ 
\beta_{\lambda_2} \frac{\partial}{\partial \lambda_2}+ 
\beta_{\lambda_3} \frac{\partial}{\partial \lambda_3}+ 
\beta_{\lambda_4} \frac{\partial}{\partial \lambda_4}\right. \nonumber\\
&&\hspace{3cm}\left. 
-\gamma_{h^0} \, h^0_c \, \frac{\partial}{\partial h^0_c}-
\gamma_{\eta^0} \, \eta^0_c \, \frac{\partial}{\partial \eta^0_c}\right) 
V_0(h^0_c, \eta^0_c) = - 2 \frac{\partial}{\partial \log{\mu^2}} V_1(h^0_c, 
\eta^0_c).
\end{eqnarray}
In terms of the tree level masses this equation is equivalent to
\begin{eqnarray}
&&\hspace{-1cm}
\left(2\,\beta_{\mu_1} - 4\,\gamma_{h^0}\, \mu_1^2\right) \, {h^0_c}^2 +
\left(2\,\beta_{\mu_2} - 4\,\gamma_{\eta^0}\, \mu_2^2\right) \, {\eta^0_c}^2 +
\left(\beta_{\lambda_1} - 4\,\gamma_{h^0}\, \lambda_1\right) \, {h^0_c}^4 
+
\left(\beta_{\lambda_2} - 4\,\gamma_{\eta^0}\, \lambda_2\right)\,{\eta^0_c}^4
\nonumber\\
&&+
\left(\beta_{\lambda_3} - 2\,\left(\gamma_{h^0}+\gamma_{\eta^0}\right)\,
\lambda_3 \right)\,{h^0_c}^2\,{\eta^0_c}^2  
- 2 \, \left(\beta_{\lambda_4}-\left(2\,\gamma_{h^0} + \gamma_{\eta^0}\right)
\,\lambda_4 \right)\,{h^0_c}^2\,{\eta^0_c}\nonumber\\
&&=\frac{1}{8 \pi^2} \left(3\,m_Z^4 + 6\, m^4_W 
-12\,m_t^4 + m^4_{\phi^0}+ 2\,m^4_{g^\pm}+
2\,m^4_{h^\pm} + m^4_{H^0} + m^4_{N^0} \right),
\end{eqnarray}
and, matching powers of fields, we can derive the beta functions:
\begin{eqnarray}
\beta_{\mu_1} &=& -\frac{2}{C_{\rm UV}} \delta {\mu_1^2} + 2 \gamma_{h^0} 
\mu_1^2,\\
\beta_{\mu_2} &=& -\frac{2}{C_{\rm UV}} \delta {\mu_2^2} + 2 \gamma_{\eta^0} 
\mu_2^2,\\
\beta_{\lambda_1} &=& \frac{2}{C_{\rm UV}} \delta {\lambda_1} + 4
\gamma_{h^0} \lambda_1,\\
\beta_{\lambda_2} &=& \frac{2}{C_{\rm UV}} \delta {\lambda_2} + 4
\gamma_{\eta^0} \lambda_2,\\
\beta_{\lambda_3} &=& \frac{2}{C_{\rm UV}} \delta {\lambda_3} + 2
\left(\gamma_{h^0}+\gamma_{\eta^0}\right) \lambda_3,\\
\beta_{\lambda_4} &=& \frac{2}{C_{\rm UV}} \delta {\lambda_4} +
\left(2\,\gamma_{h^0}+\gamma_{\eta^0}\right) \lambda_4.
\end{eqnarray}
We can now make use of the anomalous dimensions for the two neutral Higgs 
fields
\begin{eqnarray}
\gamma_{h^0} &=& \frac{1}{16 \pi^2}\,\left(3\,h_t^2 
-\frac{9}{4}\,g^2 - \frac{3}{4}\,{g'}^2\right),\\
\gamma_{\eta^0} &=& -\frac{3}{8 \pi^2}\,g^2,
\end{eqnarray}
to write down our final expressions for the one-loop beta functions:
\begin{eqnarray}
\beta_{\mu_1} &=&
\frac{1}{16 \pi^2}\,\left(6 \, \lambda_4^2 + 12 \, \lambda_1 \mu_1^2 
+ 3 \, \lambda_3 \, \mu_2^2\right)
+ \frac{1}{8 \pi^2}\,\left(3\,h_t^2 
-\frac{9}{4}\,g^2 - \frac{3}{4}\,{g'}^2\right)\,\mu_1^2,\\
\beta_{\mu_2} &=& 
\frac{1}{16 \pi^2}\,\left(4\,\lambda_4^2 + 4\, \lambda_3 \, \mu_1^2 + 10 \,
\lambda_2 \, \mu_2^2 \right)
-\frac{3}{4 \pi^2}\,g^2 \, \mu_2^2 ,\\
\beta_{\lambda_1} &=& \frac{1}{8 \pi^2}\left(\frac{9}{16}\,g^4 - 3\,{{h_t}}^4 
+12 \, \lambda_1^2 + \frac{3}{4}\,\lambda_3^2 + \frac{3}{8}\,g^2\,{g'}^2
+ \frac{3}{16}\,{g'}^4\right) \nonumber\\
&+&\frac{1}{4 \pi^2}\,\left(3\,h_t^2 
-\frac{9}{4}\,g^2 - \frac{3}{4}\,{g'}^2\right)\,\lambda_1,\\
\beta_{\lambda_2} &=& \frac{1}{8\pi^2}
\left(6\,g^4 + 11\,\lambda_2^2 + \lambda_3^2 \right)
-\frac{3}{2 \pi^2}\,g^2 \, \lambda_2,\\
\beta_{\lambda_3} &=& \frac{1}{8\pi^2}
\left(3 \, g^4 + 6\,\lambda_1 \, \lambda_3 + 5 \, \lambda_2 \, \lambda_3 
+ 2 \, \lambda_3^2\right)
+\frac{1}{8 \pi^2}\,\left(3\,h_t^2 
-\frac{33}{4}\,g^2 - \frac{3}{4}\,{g'}^2\right)\,\lambda_3,\\
\beta_{\lambda_4}&=& \frac{1}{4\pi^2}\,\lambda_4 \, 
\left(\lambda_1 + \lambda_3 \right)+\frac{3}{32 \pi^2}\,\left(4\,h_t^2 
- 7\,g^2 - \,{g'}^2\right)\,\lambda_4.
\end{eqnarray}
In the gauge and top quark 
sector the beta functions for the $U(1)$, $SU(3)$ and Yukawa 
couplings are the same as in the Standard Model, i.e.
\begin{eqnarray}
\beta_{g'} &=& \frac{41}{96 \pi^2} \, {g'}^3, \\
\beta_{g_S} &=& - \frac{7}{16 \pi^2} \, {g_S}^3,\\
\beta_{h_t} &=& \frac{1}{16 \pi^2} \, 
\left\{\frac{9}{2}\,{h_t^2}-8\,{g_S}^2 -\frac{9}{4}\,g^2-
\frac{17}{12}\,{g'}^2\right\}\,h_t.
\end{eqnarray}
The $SU(2)$ coupling is modified due to the extra Higgs triplet in the adjoint 
representation, i.e. 
\begin{eqnarray}
\beta_g &=& - \frac{5}{32 \pi^2}\,g^3.
\end{eqnarray}
Working with the tree-level effective potential with couplings
evolved using the one-loop $\beta$ and $\gamma$ functions we are able to 
resum the leading logarithms to all orders in the effective potential. 
It would be possible to include the 
next-to-leading logarithmic contributions by using the two-loop $\beta$ and 
$\gamma$ functions and including the one-loop part of the effective 
potential, see \cite{Kastening:1991,Ford:mv,Bando:1992wz}.

Let us now turn to the RG analysis. We first introduce the parameter $t$, 
related to the scale $\mu$ through  $\mu (t) = m_Z \exp{(t)}$. We
shall perform evolution starting at $t=0$.
The RG equations are coupled differential equations in the set
\begin{eqnarray}
\left\{g_s, ~g, ~g',~h_t,~\mu_1,~\mu_2,~\lambda_1,~\lambda_2,~\lambda_3
,~\lambda_4\right\}.
\end{eqnarray}
We choose rather to use the following set to define the input to the
RG equations:
\begin{eqnarray}
\left\{\alpha_s,~m_Z,~\sin^2{\theta_W},~m_t,~m_{h^\pm},~m_{H^0},~m_{N^0},~v,~
\tan{\beta},~\tan{\gamma} \right\}.
\end{eqnarray}
Within the accuracy to which we are working, the values of the 
couplings at $t=0$ can be obtained from the input set
using the appropriate tree-level expressions.

The vacuum conditions,
\begin{eqnarray}
{h^0_c} \, \mu_1^2 + \lambda_1 \, {h^0_c}^3 - \lambda_4 \, {h^0_c}\,
\eta_c^0 + \frac{1}{2}\,\lambda_3 \, {h^0_c} \, {\eta^0_c}^2 = 0, \\
{\eta^0_c} \, \mu_2^2 + \frac{1}{2}\,\lambda_3 \,{h^0_c}^2\,{\eta^0_c}
 -\frac{1}{2}\,\lambda_4\,{h^0_c}^2 + \lambda_2 \, {\eta^0_c}^3 = 0,
\end{eqnarray}
allow us to write (defining $h^0_c \equiv v$ and 
$\eta^0_c ~\equiv~ \frac{v}{2}\,\tan{\beta}$) 
\begin{eqnarray}
m_Z^2 &=&\frac{1}{4}\,v^2\,\left( g^2 + {g'}^2 \right), \\
m_W^2 &=& \frac{1}{4}\,g^2\,{v}^2\,\left(1+\tan^2{\beta}\right), \\
m_t^2 &=& \frac{1}{2}{h_t}^2\,{v}^2, \\
m_{\phi^0}^2 &=& m_{g^\pm}^2 ~=~ 0,\\
m_{h^\pm}^2 &=& v\,\lambda_4 \,\left(\cot{\beta}+\tan{\beta}\right),\\ 
m_{H^0}^2 &=& v\,\left\{2\,v\,\lambda_1 +
\left(\lambda_4 - \frac{1}{2}\,v \,\lambda_3\, \tan{\beta}\right)
\,\tan{\gamma}\right\}, \\
m_{N^0}^2 &=& v\, \lambda_4\,\left(\cot{\beta}-\tan{\gamma}\right)
+\frac{1}{2}\,v^2\,\tan{\beta}\left(\lambda_2\,\tan{\beta}+
\lambda_3\, \tan{\gamma}\right), \\
\tan{(2\,\gamma)} &=& \frac{2\,\tan{\beta}\,\left(-2\,\lambda_4 + 
v\,\lambda_3\,\tan{\beta}\right)}{2\,\lambda_4 - 4 \, v \, \lambda_1\,
\tan{\beta}+v\,\lambda_2\,\tan^3{\beta}}.
\end{eqnarray}
Inverting these relations we can thus fix the $t=0$ boundary
conditions for the subsequent evolution:
\begin{eqnarray}
g_s &\equiv& \sqrt{4 \pi \alpha_s (m_Z)} ~\simeq~ 1.22, \\
v &\equiv& \frac{1}{2^{1/4}\sqrt{G_{\rm Fermi}}} ~\simeq~ 246 {\rm ~GeV},\\
g' &\equiv& g\,\tan{\theta_W} ~\simeq~ 0.35,\\  
g &\equiv& 2\,\frac{m_Z}{v}\,\cos{\theta_W} ~\simeq~ 0.65,\\
h_t &\equiv& \sqrt{2}\,\frac{m_t}{v} ~\simeq~ 1.01,\\
\lambda_1 &=& \frac{1}{2\,v^2}\,\left(m_{H^0}^2\,{\cos^2{\gamma}} + 
    m_{N^0}^2\,{\sin^2{\gamma}} \right), \label{lam1} \\
\lambda_2 &=& -\frac{1}{v^2}\,\left\{m_{h^\pm}^2 - m_{H^0}^2 - 
        m_{N^0}^2 + m_{h^\pm}^2\,\cos (2\,\beta ) 
+   \left( m_{H^0}^2 - m_{N^0}^2 \right)\,\cos (2\,\gamma )
 \right\} \,{\cot^2{\beta}},\\ 
\lambda_3 &=& 
\frac{1 }{v^2}\,\cot{\beta}\,\left\{ m_{h^\pm}^2\,\sin(2\,\beta ) + 
      \left(-m_{H^0}^2 + m_{N^0}^2 \right) \,
       \sin (2\,\gamma ) \right\}, \label{lam3} \\
\lambda_4 &=& \frac{1}{v}\,m_{h^\pm}^2\,\cos{\beta}\,\sin{\beta},\\
\mu_1^2 &=& \frac{1}{8} \, \left\{-4\,m_{H^0}^2\,{\cos^2{\gamma}} + 
    2\,m_{h^\pm}^2\,{\sin^2{\beta}} - 
    4\,m_{N^0}^2\,{\sin^2{\gamma}} \right. \nonumber\\
&+&\left.\left( m_{H^0}^2 - m_{N^0}^2 \right) \,
     \sin (2\,\gamma )\,\tan{\beta}\right\},\\
\mu_2^2 &=& 
\frac{1}{4}\,\left\{m_{h^\pm}^2 - m_{H^0}^2 - m_{N^0}^2 + 
    m_{h^\pm}^2\,\cos (2\,\beta ) \right. \nonumber\\
&+& \left. \left( m_{H^0}^2 - m_{N^0}^2 \right) \,
     \left( \cos (2\,\gamma ) + 2\,\cot{\beta}\,\sin (2\,\gamma ) \right)
\right\}. 
\end{eqnarray}

To ensure that the system remains in a local minimum we impose the condition 
that the squared masses should remain positive, i.e.
\begin{eqnarray}
&&\lambda_4 > 0,\\ 
&&2\,v\,\lambda_1 +
\left(\lambda_4 - \frac{1}{2}\,v \,\lambda_3\, \tan{\beta}\right)
\,\tan{\gamma} > 0, \\
&&\lambda_4\,\left(\cot{\beta}-\tan{\gamma}\right)
+\frac{1}{2}\,v\,\tan{\beta}\left(\lambda_2\,\tan{\beta}+
\lambda_3\, \tan{\gamma}\right) > 0.
\end{eqnarray}
We impose the further requirement that the couplings remain
perturbative. In particular we insist that 
$|\lambda_i (t)| < 4 \pi$ for $i={1,2,3}$ and $|\lambda_4| < 4 \pi v$.
We run the evolution from $t = 0$ to $t_{\rm max}= \log{(\Lambda/m_Z)}$, with 
$\Lambda = 1$ TeV.

\section{Results in the non-decoupling regime}

In this section we present our results of the Higgs mass bounds in the
regime where the triplet Higgs cannot be arbitrarily heavy. As we shall
see in the next section, decoupling of the triplet occurs when both
mixing angles and their sum $(\beta+\gamma)$ tend to zero and in this
case, the triplet decouples from the doublet and can be arbitrarily 
heavy.

\begin{figure} 
\centerline{\psfig{file=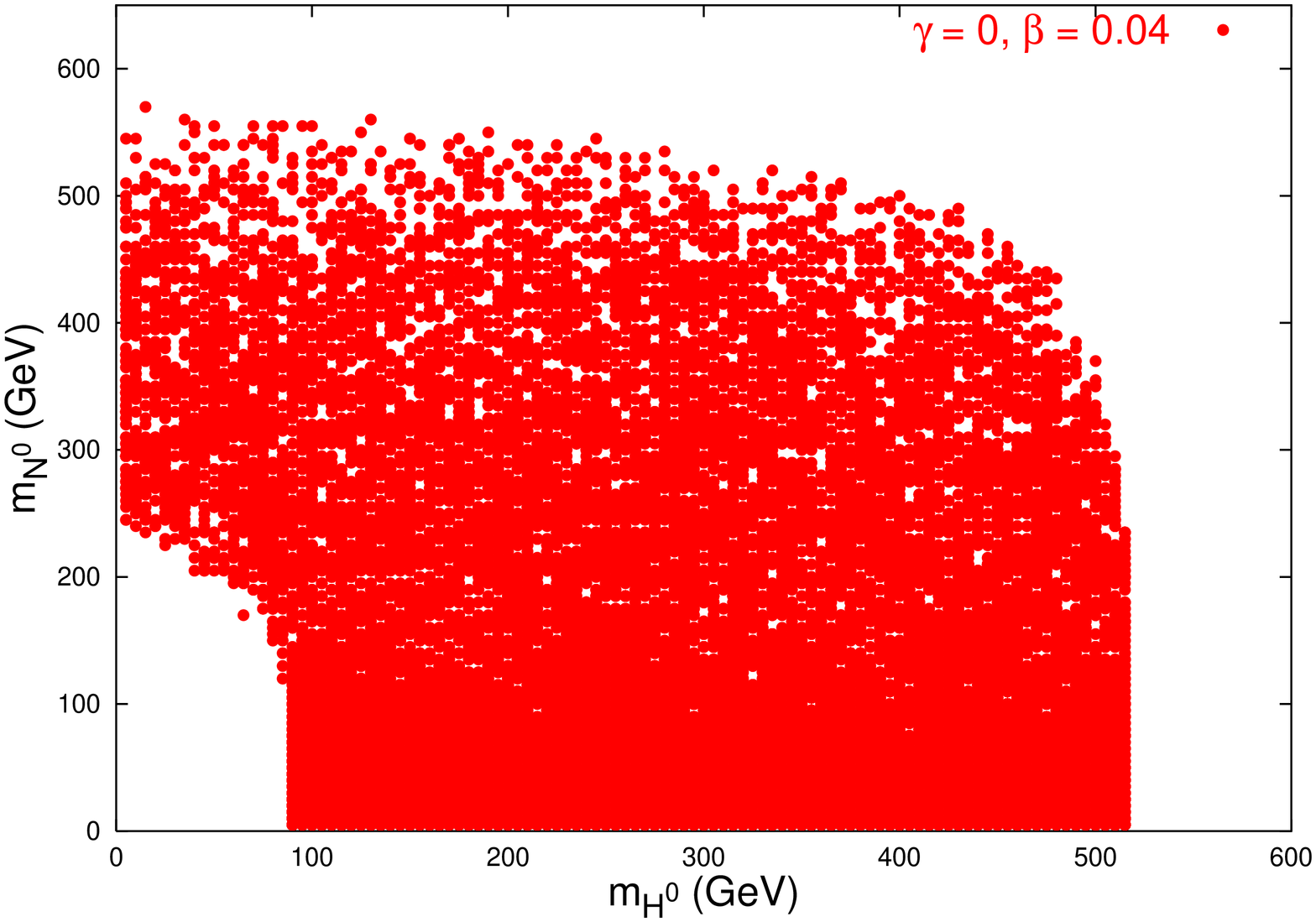,width=7.5cm,angle=-90}}
\centerline{\psfig{file=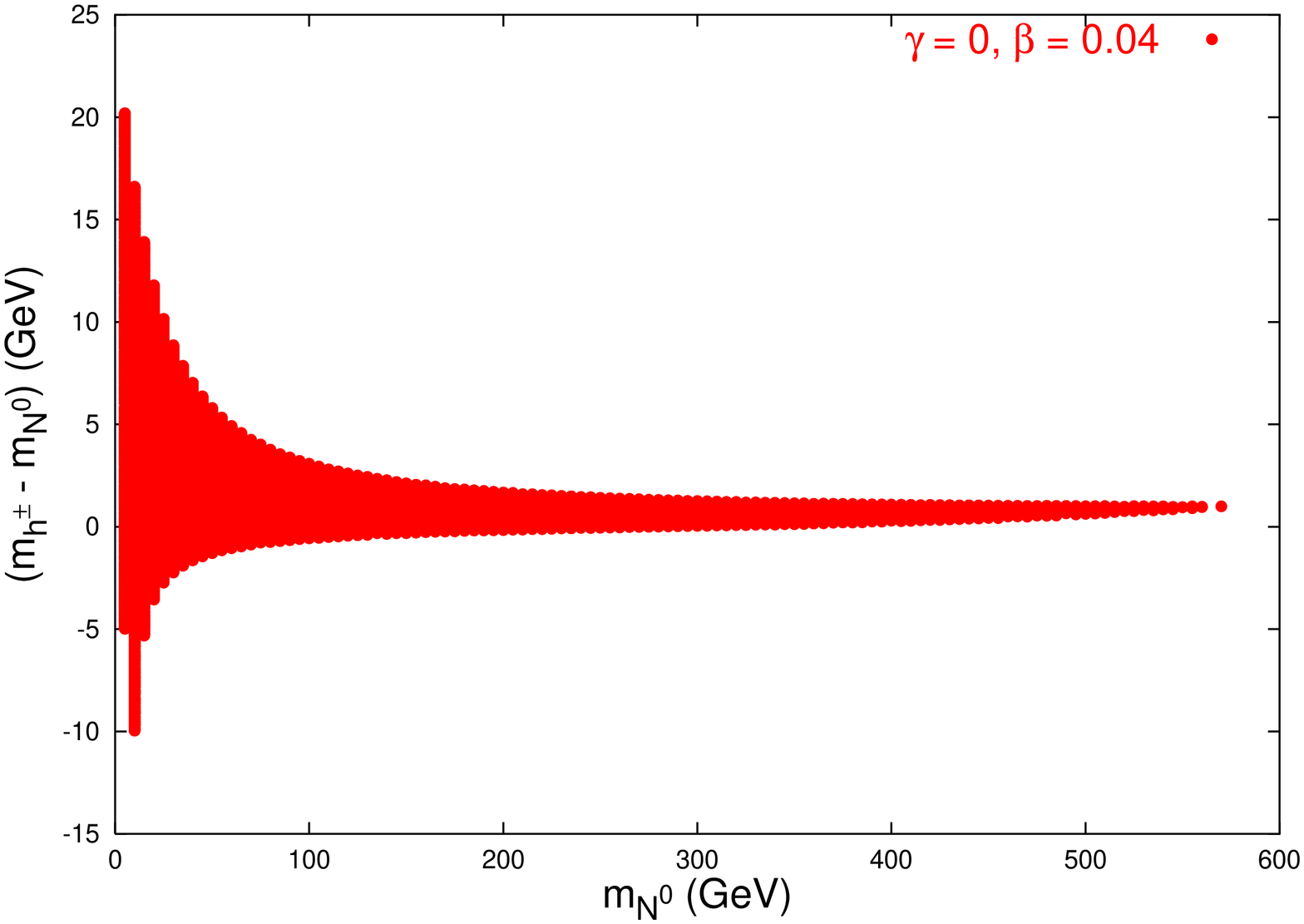,width=7.5cm,angle=-90}}
\caption{Allowed values of scalar masses for $\gamma=0$}
\label{B004G0}
\end{figure}

We are free to choose the 3 scalar masses and the 2 mixing angles at
$t=0$. In Figure \ref{B004G0} we show the range 
of Higgs masses allowed when there is no mixing in the neutral 
Higgs sector, $\gamma = 0$, for a value of $\beta = 0.04$. Such a value is
towards the upper end of the range allowed by the precision data and
is interesting because it allows a rather heavy lightest Higgs 
(e.g. for $\beta = 0.04,\; m_{H^0}>150$~GeV and for 
$\beta = 0.05, \; m_{H^0}>300$~GeV)~\cite{Forshaw:2001xq}.
The strong correlation between the $h^\pm$ and $N^0$ masses arises in order
that $\lambda_2$ remain perturbative ($\Delta m \sim \beta^2 v$ for
masses $\sim v$). The upper bound on the triplet Higgs masses 
($\approx 550$~GeV) comes
about from the perturbativity of $\lambda_3$ whilst that on $H^0$ 
($\approx 520$~GeV) comes
from the perturbativity of $\lambda_1$. These latter two bounds can
be estimated crudely by ignoring the evolution of the couplings directly from
equations (\ref{lam1}) and (\ref{lam3}). Evolution tightens the bounds
due to the positivity of the beta functions, especially for the $H^0$ since
$8 \pi^2 \beta_{\lambda_1} \approx 12 \lambda_1^2$.  
The hole at low masses is due to vacuum stability. 

\begin{figure} 
\centerline{\psfig{file=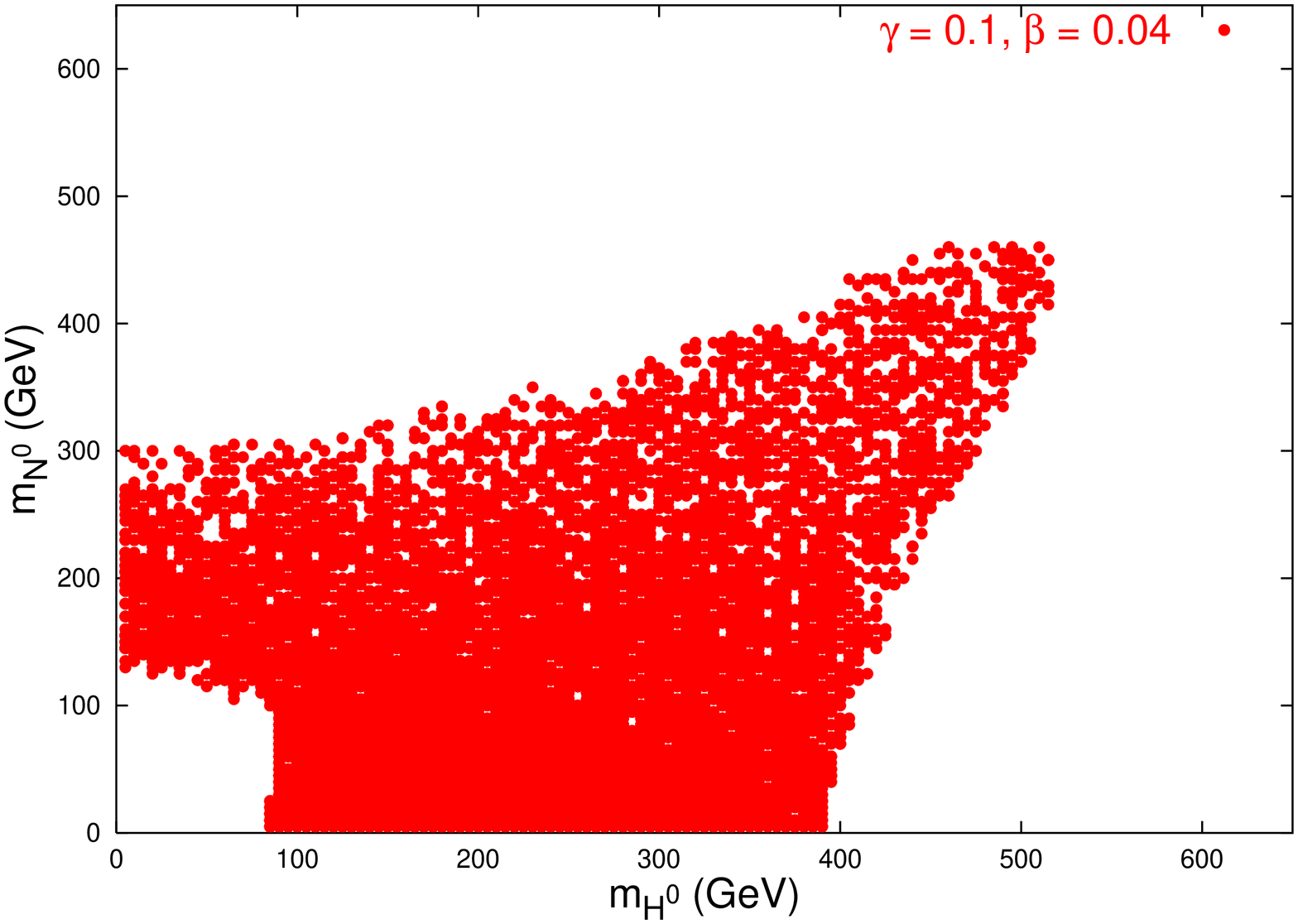,width=7.5cm,angle=-90}}
\centerline{\psfig{file=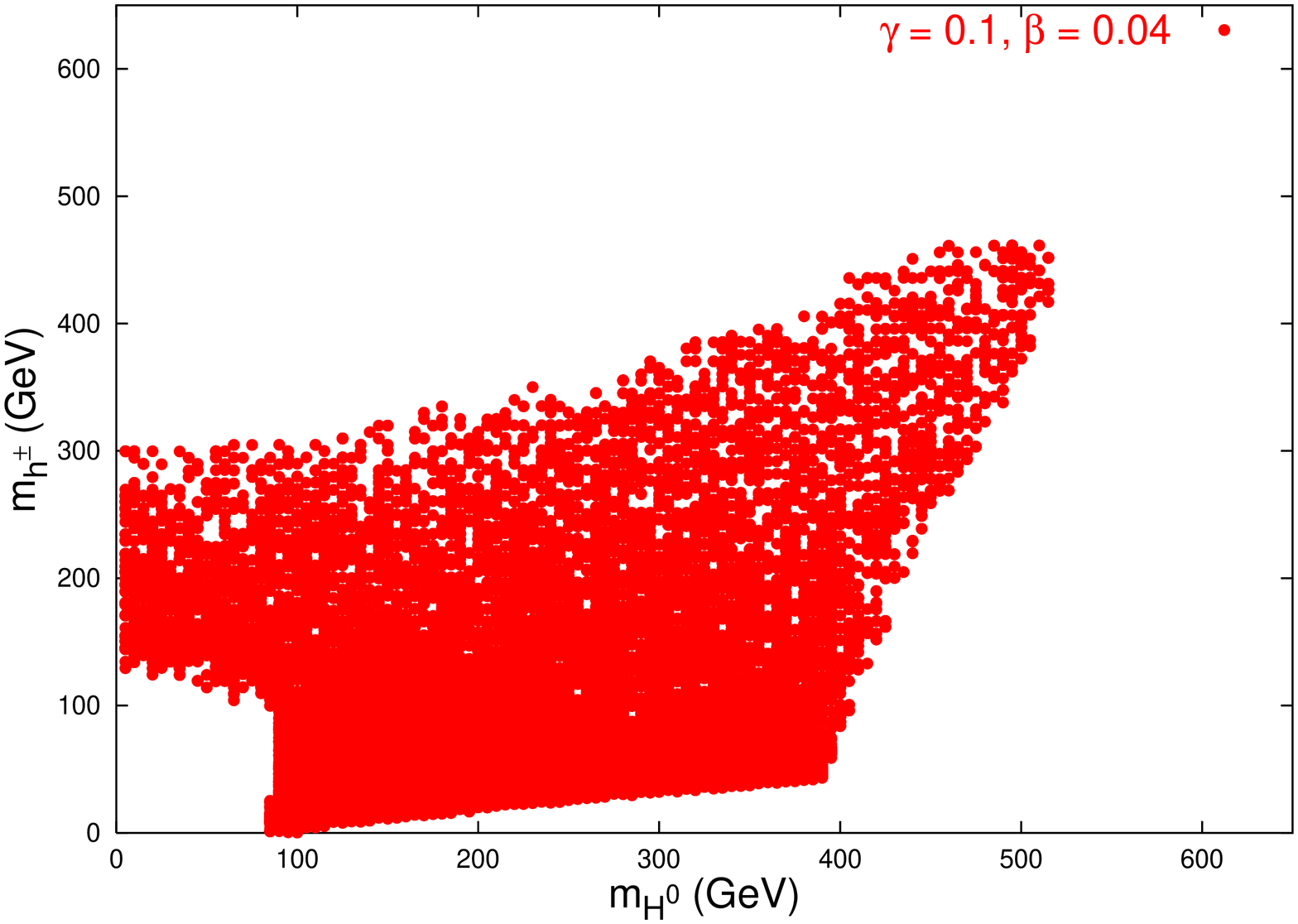,width=7.5cm,angle=-90}}
\centerline{\psfig{file=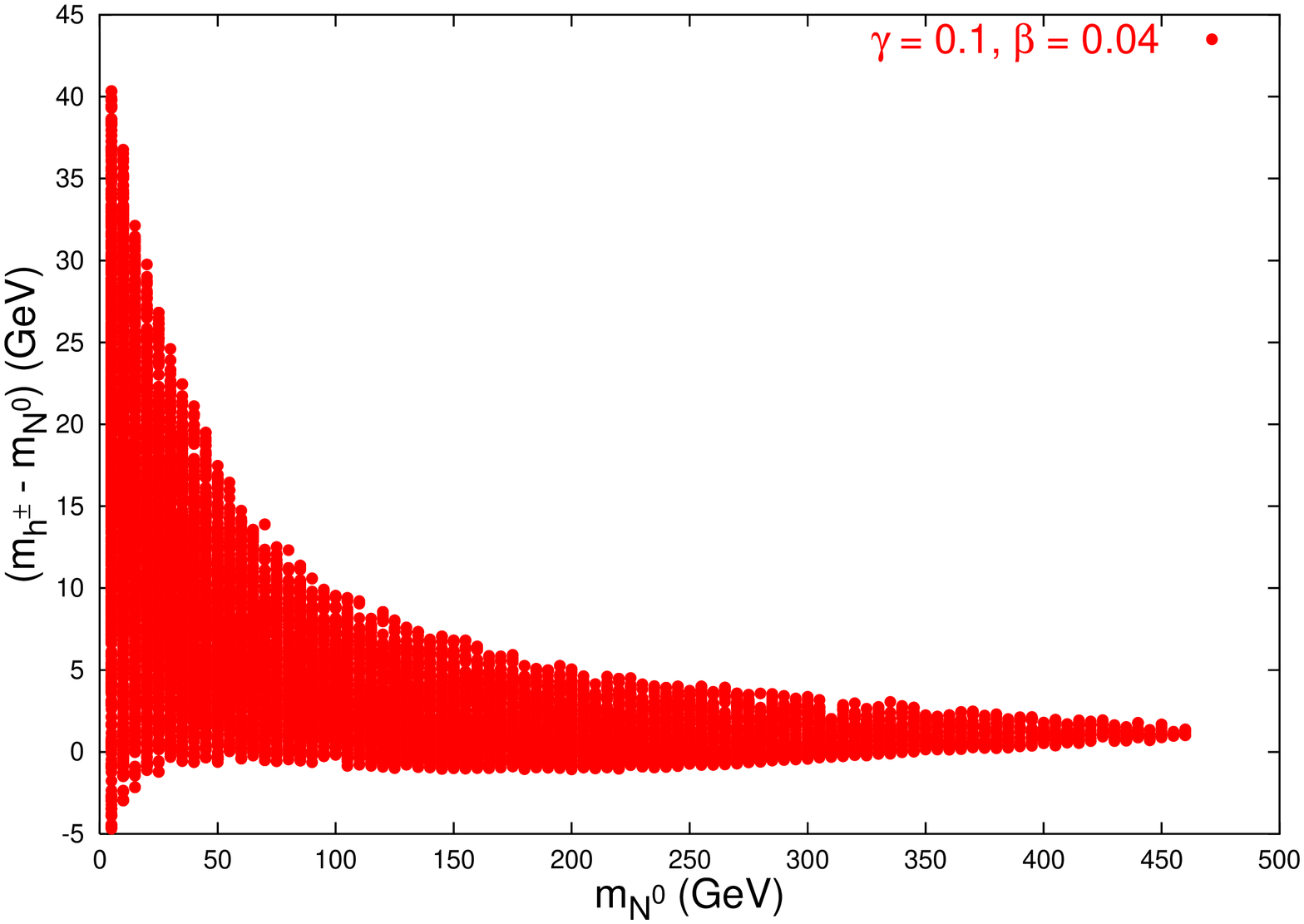,width=7.5cm,angle=-90}}
\caption{Allowed values of scalar masses for $\gamma=0.1$}
\label{B004G01}
\end{figure}

\begin{figure} 
\centerline{\psfig{file=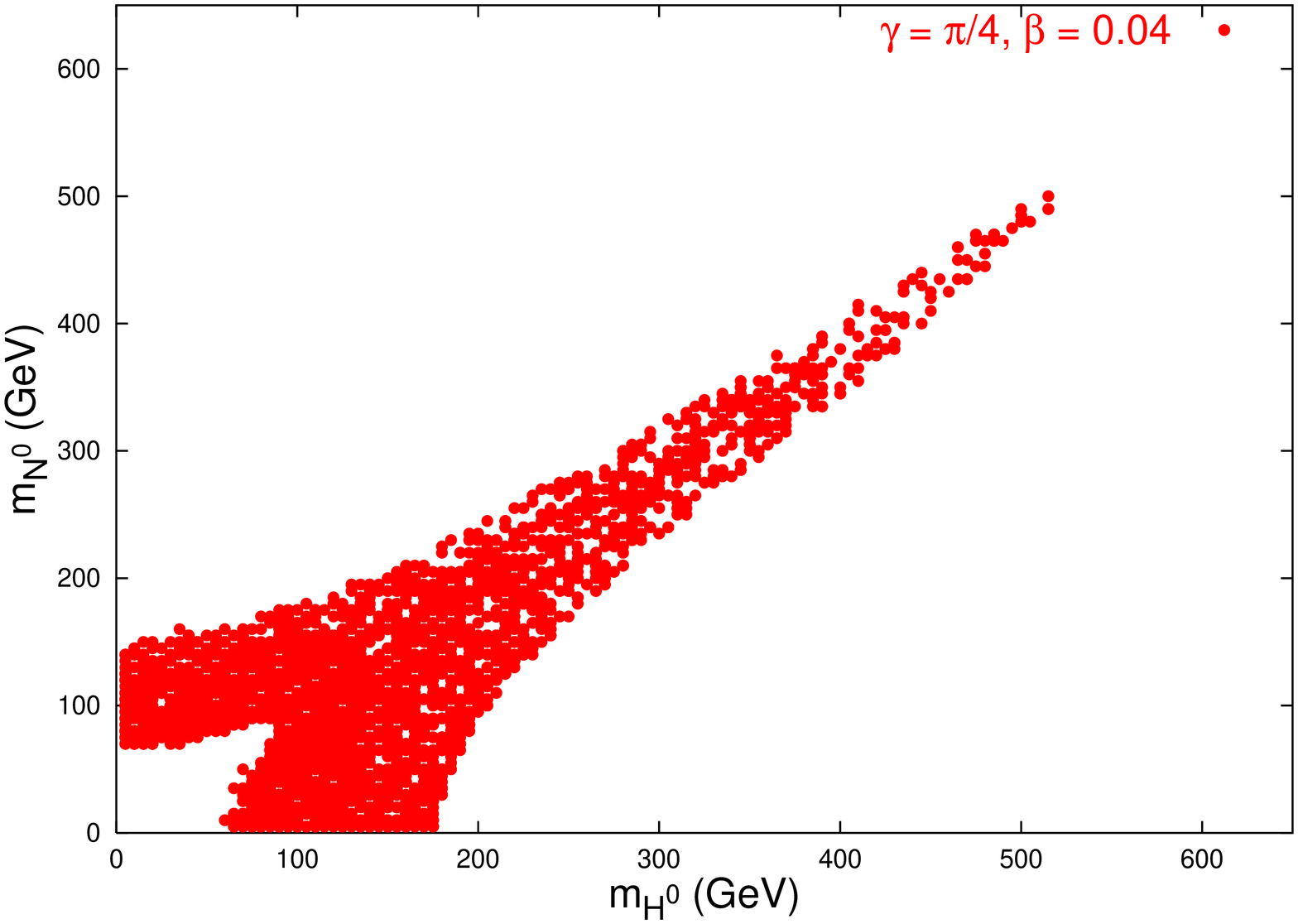,width=7.5cm,angle=-90}}
\centerline{\psfig{file=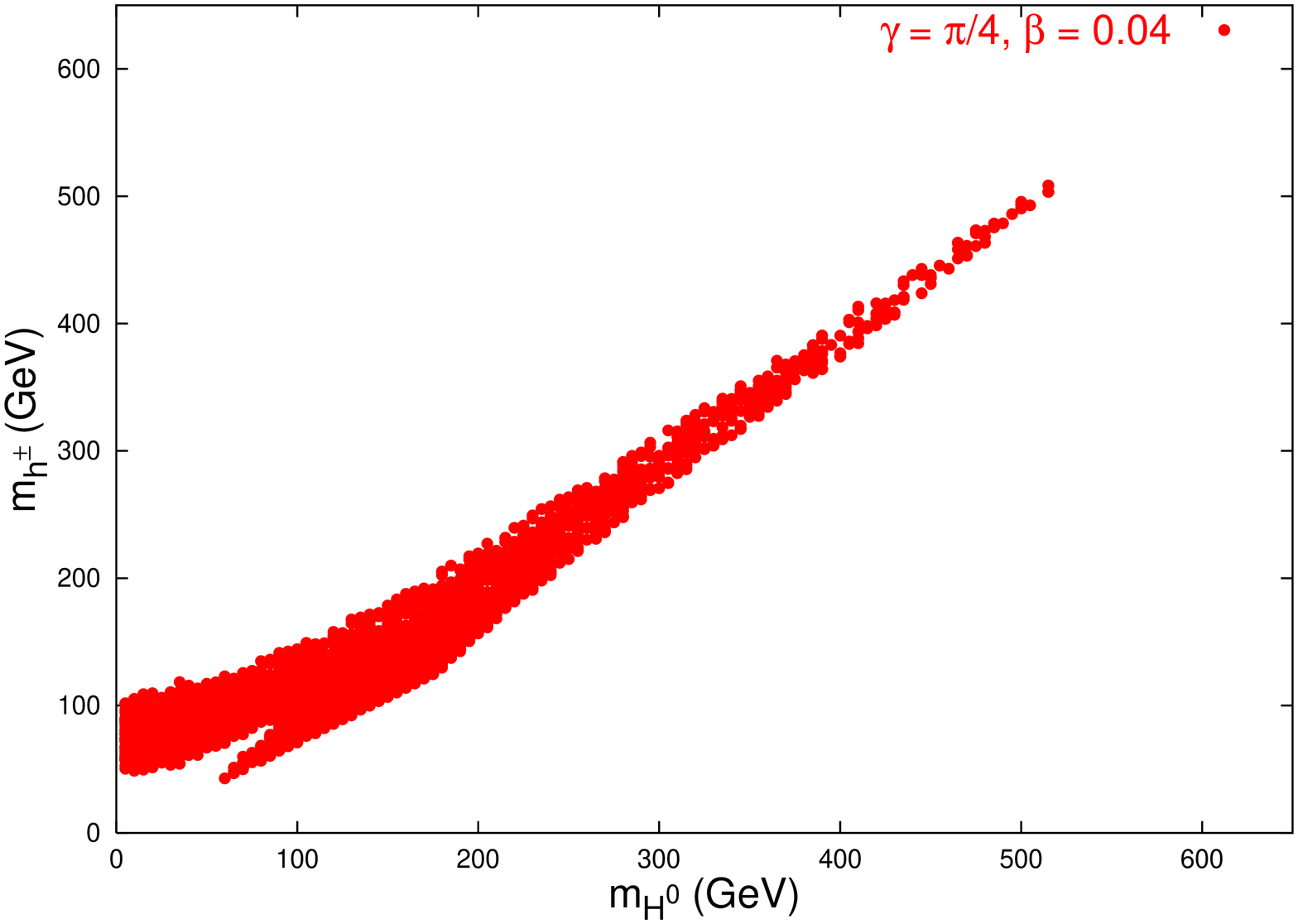,width=7.5cm,angle=-90}}
\centerline{\psfig{file=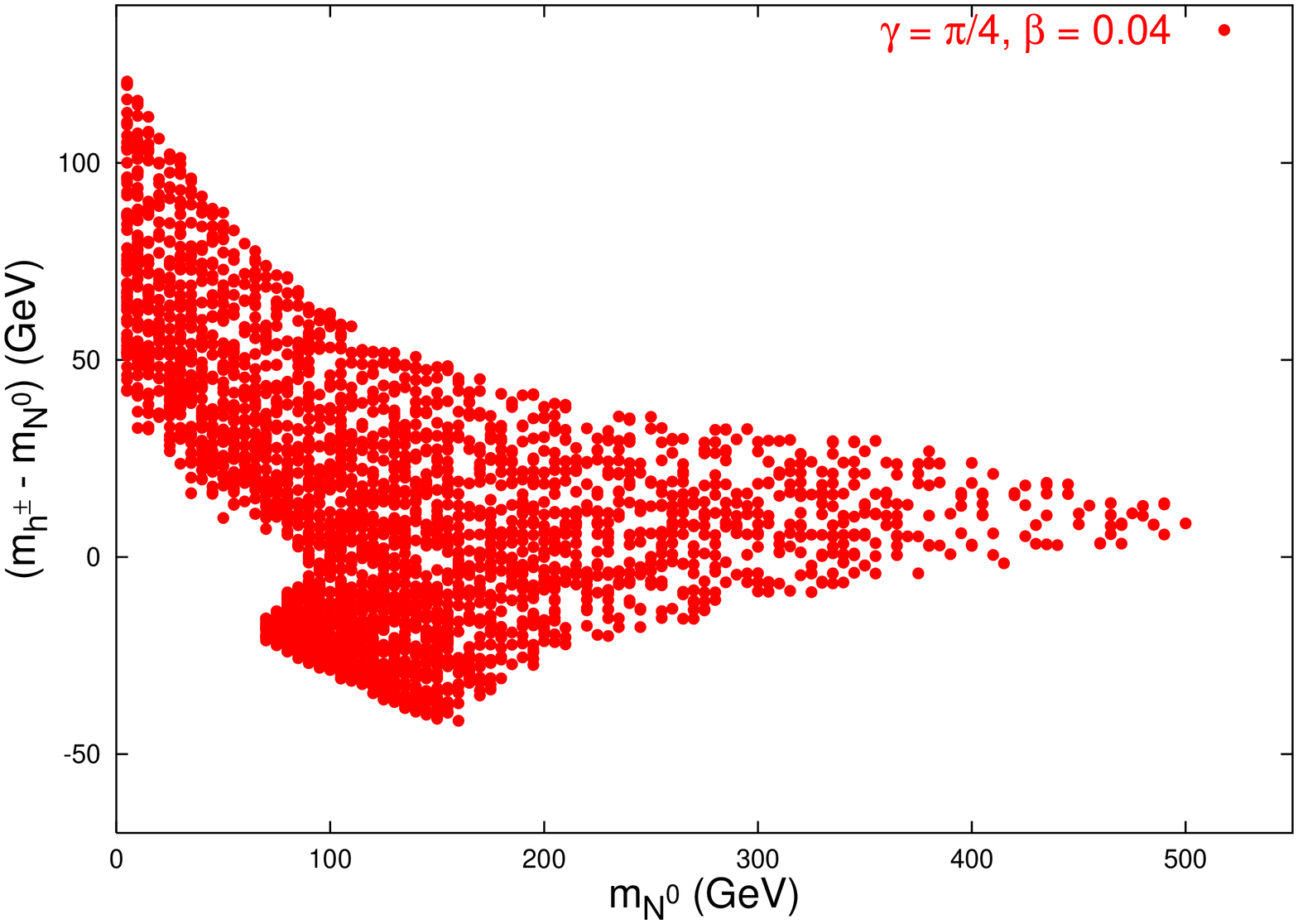,width=7.5cm,angle=-90}}
\caption{Allowed values of scalar masses for $\gamma=\pi/4$}
\label{B004GPi4}
\end{figure}

In Figure \ref{B004G01} we show the allowed regions for $\gamma = 0.1$.
The correlation of the mainly triplet Higgses is as in Figure \ref{B004G0}.
For large $m_{H^0}$ ($>450$~GeV), the upper limit on the triplet Higgs mass
arises because $\lambda_1$ becomes too large (in this region $\lambda_1
\sim \lambda_3)$. For smaller $m_{H^0}$, $\lambda_1$ is much smaller than
$\lambda_3$ and the upper bound comes from the largeness of $\lambda_3$
with the tree-level estimate being $m_{h^{\pm}}^2 < 2 \pi v^2 (\beta/\gamma)$. 
The upper limit on $m_{H^0}$ is again a consequence of
the perturbativity of $\lambda_1$, except at low $h^\pm$ masses, where it
is due to the negativity of $\lambda_3$ driving the vacuum unstable. For
very low $m_{h^\pm}$, $\lambda_2$ becoming too large is the problem.  

In Figure \ref{B004GPi4} we show the allowed regions for $\gamma = \pi/4$.
In this maximal mixing scenario one loses the distinction between doublet and
triplet Higgses and the bounds are correspondingly more democratic. The
largeness of $\tan(2 \gamma)$ can be arranged either by tuning $2 v\lambda_1
\approx \lambda_4/\beta$ or by having small enough $\lambda_1$ and 
$\lambda_4$. In the former case, all masses are approximately degenerate,
as can be seen in the plot. In the latter case, which corresponds to
light masses, the degeneracy is lifted.
The bounds for $\gamma > \pi/4$ are very similar to those for 
$(\pi/2 - \gamma)$ on interchanging the neutral Higgses $N^0$ and $H^0$. 

For $\beta < 0.04$ and small $\gamma$ (but still away from
the decoupling regime) the allowed regions are very similar to those 
for $\beta=0.04$, i.e. as in Figure \ref{B004G0}. 
For larger $\gamma$, the mass bounds are again as for larger $\beta$ 
but with the
correlation between the neutral and charged Higgs masses becoming 
even stronger than for larger $\beta$. 

We should stress that all of the previous discussion is valid
for strictly non-zero $\beta$. The situation is quite different for
$\beta = 0$. If the neutral mixing is not zero (which is required
if we are to avoid decoupling) then the vacuum conditions dictate that
$\mu_1^2 = -\lambda_1 v^2$ and $\lambda_4 = 0$ and this renders 
equation (\ref{eq:beta}) redundant. Equation
(\ref{eq:gamma}) then yields $\mu_2^2 = 2 \lambda_1 v^2 - \frac{1}{2} 
\lambda_3 v^2$ and we have complete degeneracy, i.e. 
$m^2_{H^0} = m^2_{N^0} = m^2_{h^\pm} = 2 \lambda_1 v^2$. 

\section{The decoupling limit}
So far we have worked in a regime where the triplet does not decouple
from the doublet. Clearly for $\beta = \gamma = 0$ there is no mixing
between the doublet and triplet and there is no bound on the triplet
mass. This is a special case of the more general decoupling scenario,
which occurs when $|\beta + \gamma| \ll \beta$, which we
now discuss. 

For small mixing angles, the (mainly) triplet Higgs has mass squared
$\sim \lambda_4 v / \beta$. One possible solution to the mixing angle
equations (\ref{eq:beta}) and (\ref{eq:gamma}) is that $\lambda_4 \sim
\beta v$ and any $\gamma$. In this case the triplet Higgs has mass
$\sim v$. This is the regime of the previous section. However, it is
also possible to solve the mixing angle equations with $\lambda_4 \sim v$
by keeping $\mu_2^2$ large, i.e. (\ref{eq:beta}) gives 
$\lambda_4 v = \beta \mu_2^2 \sim v^2$. In this case, equation
(\ref{eq:gamma}) forces $\beta+\gamma \approx 0$. This is the decoupling
limit in which the triplet mass lies far above the mass of the
doublet and the low energy model looks identical to the Standard Model. 

Tree level arguments on the perturbativity of $\lambda_3$ allow us to
quantify the approach to decoupling from the point of view of the
triplet Higgs mass. In particular (\ref{lam3}) dictates that, for small $\beta$
and $\gamma$,
\begin{equation}
m_{h^{\pm}}^2 \approx m_{N^0}^2 < 
\frac{2 \pi v^2 \beta + \gamma \; m_{H^0}^2}{\beta+\gamma}.
\end{equation} 
By virtue of the smallness of $\beta_{\lambda_3}$ 
this relation picks up relatively small loop corrections. This bound
clearly demonstrates decoupling. It also re-iterates the results of
the previous section, i.e. for small $\beta \gg \gamma$ the limit is as 
in Figure 1 and for small $\beta \ll \gamma$ the limit is as in Figure 2.

We remark that the pseudo-decoupling regime, where $\beta$ is not too
small, is of particular interest in that it again allows one to relax the
mass bound on the lightest Higgs coming from the precision data without
otherwise changing the physics of the Standard Model \cite{Forshaw:2001xq}.

\section{Conclusions}
We have computed the one-loop beta functions for the scalar couplings
in an extension to the Standard Model which contains an additional
real triplet Higgs. Through considerations of perturbativity of the
couplings and vacuum stability we have been able to identify the
allowed masses of the Higgs bosons in the non-decoupling regime. In the
decoupling regime, the model tends to resemble the Standard Model.

We note that the theoretical mass bounds presented here 
will of course be tightened after considering the precision electroweak
and direct search data. Such a study requires that the impact of the 
quantum corrections (to the $T$ parameter) for non-zero $\gamma$ be computed 
(they were not explored in \cite{Forshaw:2001xq}).  

As a final remark, we wish to emphasise that the near 
degeneracy of the triplet Higgs
masses (the mass splitting is naturally $\sim \beta^2 v$) ensures that,
at least for small $\gamma$, the quantum corrections to the $T$ parameter are 
negligible (the $S$ parameter vanishing since the triplet has zero 
hypercharge) \cite{Forshaw:2001xq}. As shown in \cite{Forshaw:2001xq}, 
this means that the lightest Higgs
boson can be heavy as a result of the compensation arising
from the explicit tree-level violation of custodial symmetry which the
real triplet induces. Thus it is quite
possible to be in a regime where all the Higgs bosons are heavy without
any dramatic deviation from the physics of the Standard Model.

{\bf Acknowledgements}: We would like to thank Ben Allanach, Arthur Hebecker,
Apostolos Pilaftsis and Douglas Ross for discussions. ASV acknowledges the 
support of PPARC (Postdoctoral Fellowship: PPA/P/S/1999/00446).

\end{document}